\documentclass[12pt]{article}
\usepackage{graphics}
\usepackage{color}
\usepackage{listings}
\usepackage{fullpage}
\usepackage[dvips]{lscape}
\usepackage[dvips]{graphics}
\usepackage{graphicx}
\usepackage{epsf}
\usepackage{epsfig}
\usepackage{rotating}
\usepackage{subfigure}
\usepackage{amsmath}
\usepackage{amssymb}
\usepackage{amsfonts}
\usepackage{mathrsfs}
\usepackage{latexsym}
\usepackage{authblk}
\usepackage{textpos}
\usepackage{bbm}
\usepackage{dsfont}
\usepackage{epstopdf}
\usepackage{amsthm}
\usepackage{braket}

\numberwithin{equation}{section}

\newcommand{\C}{\mathbb{C}}
\newcommand{\HH}{\mathcal{H}}
\newcommand{\A}{\mathcal{A}}
\newcommand{\Stab}{\textrm{Stab}}
\pdfoutput=1

\newcommand{\Tr}{\textrm{Tr}}
\newcommand{\Mat}{\textrm{Mat}}
\newcommand{\Orb}{\textrm{Orb}}
\newcommand{\Imm}{\textrm{Im}}
\newcommand{\Herm}{\textrm{Herm}}
\newcommand{\R}{\mathbb{R}}

\newtheorem{thrm}{Theorem}

\begin{document}

\title{\bf Locality from the Spectrum}

\author[1]{Jordan S. Cotler}
\affil[1]{\small \em  Stanford Institute for Theoretical Physics, Stanford University, Stanford CA 94305 USA}

\author[1]{Geoffrey R. Penington}

\author[1]{Daniel H. Ranard}


\maketitle

\begin{textblock*}{5cm}(12.5cm,-8cm)
\makebox{SU-ITP-17/01}
\end{textblock*}

\begin{abstract}

Essential to the description of a quantum system are its local degrees of freedom, which enable the interpretation of subsystems and dynamics in the Hilbert space.  While a choice of local tensor factorization of the Hilbert space is often implicit in the writing of a Hamiltonian or Lagrangian, the identification of local tensor factors is not intrinsic to the Hilbert space itself.  Instead, the only basis-invariant data of a Hamiltonian is its spectrum, which does not manifestly determine the local structure.  This ambiguity is highlighted by the existence of dualities, in which the same energy spectrum may describe two systems with very different local degrees of freedom.  We argue that in fact, the energy spectrum alone almost always encodes a unique description of local degrees of freedom when such a description exists, allowing one to explicitly identify local subsystems and how they interact.  As a consequence, we can almost always write a Hamiltonian in its local presentation given only its spectrum.  In special cases, multiple dual local descriptions can be extracted from a given spectrum, but generically the local description is unique.

\noindent
\let\thefootnote\relax\footnotetext{\hspace{-0.75cm}{\tt \,\,jcotler@stanford.edu, \tt geoffp@stanford.edu, \tt dranard@stanford.edu}}
\end{abstract}

\newpage
\tableofcontents
\newpage

\section{Introduction}
Quantum systems are usually described by a Hilbert space, a state vector, and a Hamiltonian.  Do these structures alone fully characterize a physical system?  Without specifying more information, like a preferred choice of basis, it is difficult to make sense of the Hamiltonian or the state.  The question of determining a preferred basis or choice of subsystems is a longstanding problem \cite{Schlosshauer1}. Here, we make progress by arguing that suitable quantum systems may be uniquely decomposed into locally interacting subsystems, given only basis-invariant data.

As a concrete example, consider the one-dimensional Ising model, with Hamiltonian
\begin{equation}\label{eq:ising}
H = J \sum_{i=1}^{n-1} \sigma^z_i \sigma^z_{i+1} + h \sum_{i=1}^n \sigma^x_i\,.
\end{equation}
The Hamiltonian clearly describes a chain of locally coupled two-level systems.  This interpretation is possible because the expression for the Hamiltonian implicitly includes a partition of the total Hilbert space into subsystems using a tensor product factorization,
\begin{equation}
\HH = \bigotimes_{i=1}^n \mathbb{C}^2\,.
\end{equation}
This choice of ``tensor product structure" (TPS) allows one to write the Hamiltonian simply in terms of local operators.  However, if one does not specify a TPS but instead writes the Hamiltonian as a large matrix in some arbitrary basis, the system becomes difficult to interpret.  Is it the one-dimensional Ising model, or is it a collection of interacting particles in three dimensions?  Up to a change of basis, different Hamiltonians are only distinguished by their energy spectra. Moreover, the only canonical choice of basis is the energy eigenbasis.\footnote{Of course, this basis is only defined up to unitaries acting within eigenspaces.}  Thus the Hamiltonian and state vector alone do not yield an obvious physical description, at least without a choice of TPS.

We therefore ask, without a preferred choice of basis, is there a natural way to decompose the Hilbert space into subsystems (i.e.\! tensor factors), knowing only the Hamiltonian?\footnote{Likewise one might ask how to most naturally divide up classical phase space into particle degrees of freedom.  For example, given four classical particles in one dimension with a contact interaction, we can equally well describe their dynamics as that of two particles in two dimensions with non-local interactions.}  In other words, do the energy eigenspaces and spectrum alone determine a natural choice of TPS? This is a question that has rarely been addressed in the literature, though it is discussed in a few papers such as \cite{piazza, tegmark}. More commonly, it has been assumed that a preferred TPS must be specified before any further progress can be made in describing the system \cite{wallace}. To even attempt finding a natural TPS, one must first specify what constitutes a natural choice.  Here, we seek a choice of subsystems such that most pairs of subsystems do not directly interact.  That is, we want the Hamiltonian to act locally with respect to the chosen TPS.  

The question of finding a natural TPS is especially relevant when one considers dualities in quantum systems.  For instance, consider the mapping 
\begin{equation}\label{eq:isingmap}
\mu_i^z = \prod_{j\leq i} \sigma_j^x\,, \quad \mu_i^x = \sigma_i^z \sigma_{i+1}^z\,, \quad \mu_n^x = \sigma_n^z\, ,
\end{equation} under which the Hamiltonian of the one-dimensional Ising model becomes
\begin{equation} \label{eq:isingdual}
H = J \sum_{i=1}^n \mu^x_i + h \sum_{i=1}^{n-1} \mu^z_i \mu^z_{i+1} \, - J \, \mu_n^x + h \, \mu_1^z
\end{equation}
where $J \, \mu_n^x$ and $h \, \mu_1^z$ are boundary terms.  The mapping demonstrates two different sets of variables, $\{\sigma_i\}$ and $\{\mu_i\}$, which define two different TPS's.  The Hamiltonian acts locally with respect to both TPS's, even though the $\{\sigma_i\}$ and $\{\mu_i\}$ operators are non-locally related to each other.  We say that the $\{\sigma_i\}$ and $\{\mu_i\}$ descriptions are ``dual," providing different local descriptions of the same Hamiltonian.

A simple argument in Section \ref{sec:existence} demonstrates that given a random Hamiltonian, there is usually no choice of TPS for which the Hamiltonian is local.  However, given a generic Hamiltonian that is local in \textit{some} TPS, we can ask whether that is the unique TPS for which the Hamiltonian is local.  In other words, given a Hamiltonian with some local description, is that local description unique? 

We present evidence that generic local Hamiltonians have unique local descriptions: that is, dualities are the exception rather than the rule. As a result, the spectrum is generically sufficient to uniquely determine a natural choice of TPS, whenever such a choice exists. We formalize a version of this statement and then prove a weaker result.  The weaker result relies on the assumption that there exists at least one example of a Hamiltonian with a unique local TPS.  (By ``local TPS," we mean a TPS for which the Hamiltonian is local.)
 
The rest of the paper is organized as follows.  In Sections 2 and 3, we formally define the notion of a TPS and what it means for a Hamiltonian to be local with respect to a particular TPS.  In Section 4, we address the main question of this paper: if a Hamiltonian is local with respect to some TPS, do we expect that TPS to be the unique TPS for which the Hamiltonian is local?  We then utilize a change in perspective analogous to the change between active and passive coordinate transformations, allowing one to re-phrase the central question as follows: given a Hamiltonian local in some fixed TPS, are there other local Hamiltonians with the same spectrum? When such Hamiltonians exist, we call them ``duals," like the two Hamiltonians of Eqn.'s~\eqref{eq:ising} and~\eqref{eq:isingdual}.  For each distinct dual, the Hamiltonian has a distinct local TPS.
 
To prove that generic local Hamiltonians have no non-trivially related duals, we break the argument in two parts.  First, we address the ``infinitesimal" version of the question: do Hamiltonians generically have duals that are infinitesimally nearby?  In Theorem 1, we address this “linearized” version of the question using a linear-algebraic argument.  Next, we address the more difficult ``global" version of the question: do Hamiltonians generically have duals related by arbitrary (non-infintesimal) transformations?  Theorem 2 addresses this question with algebraic geometry.

Both theorems discussed above rely on crucial the assumption that there exists at least one local Hamiltonian with no trivially related duals. Given such an example, we could then conclude that almost all local Hamiltonians do not have duals.  This genericity result holds for $k$-local Hamiltonians on systems of any finite size, as well as for several other notions of locality.  Restricting to the class of translation-invariant, nearest-neighbor Hamiltonians on a small number of qubits, in Section 5 we proceed to numerically find an example of such a Hamiltonian with a unique local TPS. When combined with the numerical result, the analytic result mentioned above provides an effective proof that there exists a unique local TPS for generic local Hamiltonians within this restricted class. We speculate that this conclusion extends to generic local Hamiltonians on systems of any finite size.
 
All results presented are derived for models with a finite number of finite-dimensional subsystems. Such models may be used to approximate regularized quantum field theories, although the results here are not rigorously extended to infinite dimensional Hilbert spaces. Interesting subtleties may exist for infinite-dimensional systems, both due to the possibility of continuous spectra and also due to the breakdown of analyticity, familiar from the study of phase transitions. However, we speculate that results of a similar spirit would still hold in the large-system limit.
 
In Section 6, we discuss generalizations of TPS's, needed for fermions and gauge theories. Finally, we comment on how our results frame discussions of quantum mechanics and quantum gravity.

\section{Defining Tensor Product Structures}\label{sec:tps}

Here we precisely define the notion of a tensor product structure, or TPS.  Often, one considers a Hilbert space with an explicit tensor factorization
\begin{align}
\HH_1 \otimes \HH_2 \otimes \HH_3 \otimes \cdots \,
\end{align}
where the subsystems have Hilbert spaces $\HH_i$\,.  We will usually imagine that the subsystems correspond to spatial lattice sites.  (In few-body quantum mechanics, the subsystems might correspond to distinguishable particles, whereas in many-body physics or regularized quantum field theory, the subsystems might correspond to lattice sites, momentum modes, quasiparticle modes, or some other choice.)    

Our first task is to define a TPS on an abstract Hilbert space $\HH$ that is not written as an explicit tensor product.  This formalism will lend precision to the discussion of different tensor product structures on the same Hilbert space, the topic at the heart of this paper.

Consider a map $T$ on a Hilbert space $\HH$ which is an isomorphism (unitary map)
\begin{equation}
T : \HH \to \HH_1 \otimes \HH_2 \otimes  \cdots\,.
\end{equation}
The choice of isomorphism endows $\HH$ with a notion of locality: one can then speak of local operators, subsystems, entanglement, and so on within the Hilbert space $\HH$.  For instance, we say the operator $O$ on $\HH$ is local to subsystem $i$ if $TOT^{-1}$ is local to $\HH_i$.  Similarly, the entanglement entropy of a state $\psi \in \HH$ is defined as the entanglement entropy of $T\psi$.  These notions will remain unchanged if $T$ is composed with a map $U_1 \otimes U_2 \otimes \cdots $ that acts unitarily on each subsystem.  We therefore define a TPS as follows: \\

\noindent \textbf{Definition (TPS):} A TPS $\mathcal{T}$ of Hilbert space $\HH$ is an equivalence class of isomorphisms $T : \HH \to \HH_1 \otimes \HH_2 \otimes  \cdots$, where $T_1 \sim T_2$ whenever $T_1 T_2^{-1}$ may be written as a product of local unitaries $ U_1 \otimes U_2 \otimes \cdots $ and permutations of subsystems.
\\

To avoid confusion, note that the usage of ``local" in the phrase``local unitary" is distinct from its usage in ``local Hamiltonian."  Local unitaries are products of unitaries acting on single tensor factors, while local Hamiltonians are sums of operators acting on small subsets of tensor factors.

Another equivalent and useful way to define a TPS  involves observables rather than states.  In short, a TPS naturally defines subalgebras of observables local to each subsystem, but we can turn this data around and use the subalgebras to define the TPS.  This perspective was developed by \cite{ZLL} and was studied in the context of quantum gravity in \cite{giddings1,giddings2}.  Let us collect the local observables as a set of mutually commuting subalgebras $\A_i \in L(\HH)$, where $L(\HH)$ denotes the algebra of operators on $\HH$, and $\A_i$ denotes the algebra of operators of the form
\begin{equation*}
\mathds{1} \otimes \cdots \otimes \mathds{1} \otimes O_i \otimes \mathds{1} \otimes \dots \otimes \mathds{1}\,,
\end{equation*}
i.e.\ operators that act as the identity on all subsystems except $i$. 

With this motivation, we can equivalently define a TPS on $\HH$ as a collection of of subalgebras $\{\A_i\}$, $\A_i \in L(\HH) $, such that the following hold:
\begin{enumerate}
\item  The $\A_i$ mutually commute, $[\A_i, \A_j]=0$ for $i \neq j$.
\item  The $\A_i$ are independent, $\A_i \cap \A_j = \mathds{1}$.
\item The $\A_i$ generate the whole algebra of observables, $\bigvee_i \A_i = L(\HH)$.
\end{enumerate}
The above definition is equivalent to the first definition, because a choice of subalgebras subject to the above conditions will uniquely determine an equivalence class of isomorphisms $\mathcal{T}$ that give rise to the corresponding subalgebras.  (The equivalence of definitions essentially follows from the result stated in \cite{ZLL}.)  Since one can specify a TPS either by an equivalence class $\mathcal{T}$ or a collection of subalgebras $\{\mathcal{A}_i\}$, we will switch between these notations freely.

For a Hilbert space $\HH$ without additional structure like a Hamiltonian, no choice of TPS is more interesting or meaningful than another.  That is, even though two choices $\{\A_i\}$ and $\{\A'_i\}$ may differ, neither choice is significant on its own because the states and operators of $\HH$ have no identifying structure to begin with.  However, we will be interested in a Hilbert space equipped with a Hamiltonian $H$, and this additional structure \textit{does} distinguish certain states and operators, namely the operator $H$ and its eigenvectors.  The choice of TPS then acquires more meaning; for instance, the ground state might be entangled with respect to one TPS but separable in another.  

On the other hand, certain TPS's will be effectively equivalent with respect to a given Hamiltonian.  One may think of the operator $T H T^{-1}$ on $\bigotimes_i \mathcal{H}_i$ as an expression of $H$ with respect to TPS $\mathcal{T}$.  Moreover, two operators on $\bigotimes_i \mathcal{H}_i$ describe physically equivalent systems if they are the same up to conjugation by local unitaries, permutation of subsystems, and transposition. That is, conjugation by local unitaries is merely a re-labeling of the basis within tensor factors, while permutation of subsystems is merely a re-labeling of how subsystems are labeled. And transposition of the entire Hamiltonian,\footnote{Note that although transposition is not itself a unitary map, for any given $H$, there exists a unitary operator $U$ such that $U(THT^{-1})U^\dagger = (THT^{-1})^T$ since transposition preserves the spectrum.} namely $THT^{-1} \mapsto (THT^{-1})^T=(THT^{-1})^*$, only corresponds to complex conjugation, or a relabeling of $i \mapsto -i$.

We therefore say that two TPS's $\mathcal{T}_1$ and $\mathcal{T}_2$ are equivalent \textit{with respect to the Hamiltonian H} if the operators $T_1H T_1^{-1}$ and $T_2 H T_2^{-1}$ on $\bigotimes_i \mathcal{H}_i$ are the same up to conjugation by local unitaries, permutation of subsystems, and transposition.  

The above may be encapsulated by the following definition.  Although this definition may appear unnecessarily formal, we hope it will eliminate any vagueness of the discussion, and the definition will not require careful parsing to follow the paper in general.
\\ \\
\noindent \textbf{Definition (Equivalence of TPS):} Two choices $(\HH,H,\{\A_i\})$ and $(\HH',H',\{\A'_i\})$  of Hilbert space, Hamiltonian, and TPS are equivalent when there exists a unitary $U : \HH \to \HH'$ such that $H'=U^{-1}HU$ or\footnote{Rather than allowing $H'=U^{-1}H^TU$ in addition to $H'=U^{-1}HU$, one could allow $U$ to be an anti-automorphism of the $C^*$-algebra of observables.  Automorphisms of the algebra of observables are given by unitary conjugation, while anti-automorphisms (multiplication-reversing homomorphisms) are given by the composition of unitary conjugation and transposition.} $H'=U^{-1}H^TU$, and such that $\A'_{j_i}=U^{-1}\A_i U$, where $j_1,...,j_n$ is some permutation of $i=1,...,n$.  Alternatively, $(\HH,H,\mathcal{T})$ and $(\HH',H',\mathcal{T}')$ are equivalent when there exists a unitary $U : \HH \to \HH'$ such that $H'=U^{-1}HU$ or $H'=U^{-1}H^TU$,  and such that $[T]=[T'U]$. 
\\ \\
\indent We will be most interested in considering the same Hilbert space and Hamiltonian with different TPS's, given by $(\HH,H,\mathcal{T})$ and $(\HH,H,\mathcal{T}')$.  Rather than talk about two different TPS's $\mathcal{T},\mathcal{T}'$ for the same Hamiltonian $H$, we can often simplify the discussion by talking about two different Hamiltonians $THT^{-1}, T'HT'^{-1}$ on the space $\bigotimes_i \HH_i$ with fixed TPS.  Both perspectives are equivalent.  This observation will be important and bears repeating.
\\

\noindent \textbf{Observation:}  It is equivalent to consider either perspective:
\begin{enumerate}
\item  A Hilbert space $\mathcal{H}$ with fixed Hamiltonian $H$ and varying choice of TPS $\mathcal{T}$ or $\mathcal{T}'$, or
\item  A Hilbert space $\bigotimes_i \HH_i$ with fixed TPS and unitarily varying choice of Hamiltonians $THT^{-1}$ or $T'HT'^{-1}$ with the same spectrum.  
\end{enumerate}

\noindent For some fixed Hamiltonian $H$, questions about the existence of a TPS in which $H$ is local may then be translated into questions about the existence of local Hamiltonians with the same spectrum as $H$.  

The notion of duality can also be expressed in either of these perspectives: \\

\noindent \textbf{Definition (Dual):}  From the first perspective above, we say that two TPS's are dual if the given Hamiltonian is local in both TPS's and if also the TPS's are inequivalent with respect to that Hamiltonian.  From the second perspective, we say that two Hamiltonians are dual if they are local, have the same spectrum, and cannot be related by local unitaries, permutations of subsystems, and transposition.
\\

The results of Sections \ref{sec:main} and \ref{sec:numerics} will largely be cast in the second perspective, i.e. as statements about the existence of different local Hamiltonians with the same spectrum.  However, the results may always be re-cast in the first perspective, as results about the existence of different local TPS for the same Hamiltonian.  For instance, we can either say that generic local Hamiltonians uniquely determine a local TPS, or we can say that the spectrum of a generic local Hamiltonian will uniquely determine the Hamiltonian.

\section{Defining Locality}\label{sec:locality}

Given a Hilbert space $\HH$ and Hamiltonian $H$, what qualitatively distinguishes different choices of TPS?  Most broadly, we might ask whether some TPS's yield simpler, more meaningful, or more calculationally tractable descriptions of a system.  More specifically, we are interested in TPS's for which the Hamiltonian appears local, in the sense that it only exhibits interactions among certain collections of subsystems.

We should clarify what it means for a Hamiltonian to include an interaction among a given collection of tensor factors.  In general, one can write a Hamiltonian on $n$ qudits (i.e., $n$ systems of local dimension $d$) as 
\begin{equation}
\label{locHam1}
H = a_0 \mathds{1} + \sum_{i=1}^n \sum_{\alpha=1}^{d^2-1} a_\alpha^i O^\alpha_i + \sum_{i<j} \sum_{\alpha,\beta=1}^{d^2-1} a_{\alpha\beta}^{ij} O^\alpha_i O^\beta_j +\sum_{i<j<k}\, \sum_{\alpha,\beta,\gamma=1}^{d^2-1} a_{\alpha\beta\gamma}^{ijk} O^\alpha_i O^\beta_j  O^\gamma_k + \cdots
\end{equation}
where the operators $O^\alpha_i$ for $\alpha=1,...,d^2-1$ form an orthogonal basis for single-qudit operators on site $i$.  This decomposition is unique, up to the choice of basis $O^\alpha_i$.  The terms $O^\alpha_i O^\beta_j  O^\gamma_k$, for instance, are considered as interactions between qudits $i$, $j$, and $k$.  The space of operators $L(\HH)$ decomposes into orthogonal sectors, with one sector for each combination of subsystems, and we say that $H$ contains an interaction among some subset of qudits if $H$ has a nonzero component in the corresponding sector.

Qualitatively, we say that a Hamiltonian is local when relatively few combinations of subsystems are interacting.  (Note that we use ``local operator" to refer to an operator that is local to a single subsystem or collection of subsystems, while we use ``local Hamiltonian" to refer to a sum of such operators.)  For instance, the Ising model only exhibits nearest-neighbor interactions, as do lattice-regularized quantum field theories without higher derivatives. Meanwhile, other models like spin glasses and matrix models exhibit interactions among all particles, but only in groups of fixed size, like two or four.  Likewise, non-relativistic electrons have only pairwise interactions, if each electron is treated as a subsystem.\footnote{The formalism developed here will not directly treat fermions; see Section \ref{sec:gauge}.}

To incorporate all these notions of locality, one can use a hypergraph.  First, note that an ordinary graph can be thought of as a collection of vertices $V$, along with a collection of edges $E = \{E_i\}$, where each edge is written as a pair of vertices, $E_i = \{v, v'\}$ for $v,v' \in V$.  We emphasize that each $E_i$, called an ``edge," is a two element subset of $V$.  A hypergraph is like an ordinary graph with a set $V$ of vertices, but the ``edges" $E_i \subset V$ may contain more than two vertices.  For convenience, we subsequently refer to a hypergraph as a graph.  Given a fixed TPS and Hamiltonian, the associated ``interaction graph" has vertices corresponding to the subsystems and has (hyper-)edges corresponding to every combination of subsystems that interact under the Hamiltonian.  We say that the Hamiltonian is local with respect to some graph $G$ if its interaction graph is a subgraph $G$.

Given a Hamiltonian $H$, different choices of TPS $\mathcal{T}$ will give rise to different operators $THT^{-1}$ on $\bigotimes_i \mathcal{H}_i$, with different associated interaction graphs.  We are interested in TPS's which give rise to interaction graphs with edges connecting only a small number of sites.  As one measure of sparsity or locality, we say that a graph is $k$-local if it has edges joining at most $k$ vertices.\footnote{This property is also important in few-body quantum mechanics. For an $N$-particle first quantized quantum system, $k$-locality with respect to particle subsystems means that the particles only have $k$-body interactions.}  Likewise, we call a Hamiltonian $k$-local with respect to some TPS if the associated graph is $k$-local, and we will also refer to the TPS as $k$-local.

While the study of generic $k$-local Hamiltonians is important,\footnote{In fact, as mentioned above, an $N$-particle first quantized quantum system with $k$-body interactions is $k$-local with respect to particle subsystems.} for example in the study of quantum circuits and black holes, generally we are interested in the stronger condition of geometric locality. This means that each site has edges connecting it to only a small number of other sites. For example, we are often interested in graphs which form a $d$-dimensional lattice with only neighboring lattice sites interacting.  Such graphs are ubiquitous since they arise in any local spin system or lattice regularization of quantum field theory.  We might even make further constraints such as requiring that the Hamiltonian be translation-invariant with respect to the lattice.

All of the analytic results in this paper will be valid for generic Hamiltonians within any specified locality class, including all the classes discussed above.  Specifically, we prove results about the number of duals within a particular locality class of a generic Hamiltonian in that same locality class. For example, we can prove results about the number of translation-invariant duals of a generic translation-invariant Hamiltonian. It is harder to prove results about the number of $k$-local duals of generic translation-invariant Hamiltonians, since translation-invariant geometrically local Hamiltonians are a measure zero subspace of the larger space of $k$-local Hamiltonians. However in Section \ref{sec:numerics} we show that some of our results can be extended to such cases.

\section{Existence and Uniqueness of Local TPS} \label{sec:main}
\subsection{Existence of local TPS} \label{sec:existence}

First we ask whether a generic Hamiltonian has any $k$-local TPS. The answer is no, as will be demonstrated.  We restrict our attention to a finite-dimensional Hilbert space $\HH$, $\textrm{dim}(\HH)=N$, with Hamiltonian $H$. We ask whether there exists a TPS with $n$ subsystems such that the Hamiltonian is $k$-local.  For $n$ sufficiently larger than $k$, we will see that a $k$-local TPS exists only for a measure zero set of operators in $L(\HH)$.\footnote{We only deal with sets of measure zero, so the exact measure considered is irrelevant; only the specification of measure \textit{zero} sets is important.  One could choose equally well the Gaussian unitary ensemble, or the Lebesgue measure on the space of Hermitian matrices in some arbitrary basis, both of which have the same measure zero sets.}

Recall from the previous section that for a given Hamiltonian $H$, a choice of TPS  $\mathcal{T}$ produces a Hamiltonian $THT^{-1}$ on $\bigotimes_i \HH_i$, up to local unitaries and permutations of subsystems.  The operator $THT^{-1}$  then defines some associated interaction graph, up to relabeling of vertices.  We call the TPS $\mathcal{T}$ $k$-local if it gives rise to a $k$-local interaction graph for  $THT^{-1}$.  

Note that for any TPS $\mathcal{T}$, $H$ and $THT^{-1}$ have the same spectrum.  Conversely, if there is some operator $O$ on $\bigotimes_i \HH_i$ with the same spectrum as $H$, then there exists a TPS $T'$ such that $O=T'HT'^{-1}$.  So $H$ has a $k$-local TPS if and only if there is a $k$-local Hamiltonian on $\bigotimes_i \HH_i$ with the same spectrum.  

The above observation allows a change of perspective, as suggested at the end of Section \ref{sec:tps}.   Rather than asking whether generic Hamiltonians on an abstract Hilbert space $\HH$ have some $k$-local TPS, we can equivalently ask whether generic Hamiltonians on $\bigotimes_i \HH_i$ are isospectral to some $k$-local Hamiltonian.  A simple dimension-counting argument yields the answer.  The space of possible spectra of all Hamiltonians is $\R^N$.  Meanwhile, examining Eqn. (\ref{locHam1}), we see that the space of $k$-local Hamiltonians on  $\bigotimes_i \HH_i$ will have dimension 
\begin{equation}
s = \sum_{j=0}^k \binom{n}{j} (d^2 - 1)^j
\end{equation}
for $n$ subsystems of local dimension $d$.  Then the space of spectra of $k$-local Hamiltonians will also have dimension at most $s$.  For $s<N=d^n$, the space of all spectra of $k$-local Hamiltonians will have positive codimension in the space of all possible spectra.  So for any sufficiently large $n$ (e.g. $n \geq 10$, for $d=2$ and $k=2$), the set of Hamiltonians that are isospectral to a $k$-local Hamiltonian has measure zero.

In general, the results in this paper will apply to all Hamiltonians in some specified subspace, excluding an exceptional set of measure zero.  On the other hand, when asking questions of an approximate nature -- for instance, when asking whether a generic Hamiltonian has a TPS that is \textit{approximately} local -- the relevant question is not quite ``Does the exceptional set have measure zero?" but rather ``What is the volume of an $\epsilon$-neighborhood of the exceptional set?"   Such questions are more difficult to tackle directly, requiring analysis to augment the linear algebra and algebraic geometry used in this paper.  However, the exceptional sets in question not only have measure zero but also have a codimension that is exponential in the system size, perhaps suggesting that the desired results about $\epsilon$-neighborhoods would hold.

\subsection{Uniqueness of local TPS}

Now we ask, given a Hamiltonian $H$ with some $k$-local TPS $\mathcal{T}$, is $\mathcal{T}$ the unique $k$-local TPS, up to equivalence in the sense of Section \ref{sec:tps} above?  We again follow the strategy of reformulating the question on the space $\bigotimes_i \HH_i$, using the observation at the end of Section \ref{sec:tps}.  Recall that two $k$-local Hamitonians on $\bigotimes_i \HH_i$ are called dual if they are isospectral and are not related by local unitaries, permutations of subsystems, or transposition.

Now we can reformulate the question of whether a $k$-local TPS for a Hamiltonian is generically unique. The question becomes, does a generic $k$-local Hamiltonian $H$ on $\bigotimes_i \HH_i$ have any duals?  In other words, can one generically recover a local Hamiltonian from its spectrum?  Posed in the latter terms, the question may be interesting for independent reasons.  However, we are motivated by the original question, asking whether a $k$-local TPS for a Hamiltonian is generically unique. 

One's initial intuition may suggest that a $k$-local Hamiltonian may indeed be recovered from its spectrum.  This intuition is due to dimension counting: a $k$-local Hamiltonian is specified by a number of parameters polynomial in $n$, while the number of eigenvalues is exponential in $n$.  In the previous section, this dimension counting was used to rigorously demonstrate that generic Hamiltonians have no $k$-local TPS.  However, the argument here is less immediate.   While the spectrum has more parameters than the Hamiltonian, this fact alone does not prevent generic Hamiltonians from having duals.  For instance, imagine that we slightly modified the question, instead defining two $k$-local Hamiltonians to be dual whenever they are not related by local unitaries or permutations, failing to include the possibility of transposition.  Then we would discover that all Hamiltonians (except real-symmetric ones) have at least one dual, given by their transpose. Thus it is not immediately obvious that most $k$-local Hamiltonians do not have duals.

However, transposition, just like local unitaries and permutations, is a linear map on the space of Hamiltonians, and it preserves the subspace of local Hamiltonians. A simple check, after dimension-counting, is to ask whether there are any \textit{other} linear spectrum-preserving maps that preserve this subspace. It turns out that no such maps exist. All linear maps that preserve eigenvalues are generated by transposition and unitaries, and since we already know that taking the transpose preserves locality, we only have to worry about whether other unitaries preserve the subspace. Indeed, the only unitaries that preserve the subspace are compositions of single-site unitaries and permutations.
 
We therefore argue the following lemma: if a unitary preserves the space of $k$-local Hamiltonians when acting by conjugation, the unitary must be generated by 1-local (single-site) operators and permutations of sites.  While this lemma is not central to the main arguments of the paper, it may be illuminating.

To begin, we argue an infintesimal, or ``linearized," version of the above: given a Hermitian operator $V$ such that the map $C_V : X \mapsto i[V,X]$ preserves the subspace of $k$-local Hamiltonians, $V$ must be a sum of single-site operators.  Assume $V$ is an operator such that $C_V(X)=i[V,X]$ is $k$-local for any $k$-local $X$.  Decomposing $V$ as in Equation \ref{locHam1}, let $k_{max}$ be the maximum number of interacting sites, and let $V_{max} \neq 0$ be the sum of terms in $V$, involving all sites on some subset $S$ of size $k_{max}$.  Suppose $X$ is a $k'$-site operator for $k' \leq k$ that intersects $S$ at a single site and that does not commute with $V$.  Then $i[X,V_{max}]$ contains a term with interaction size $k_{max}+k'-1$, and $i[X,V]$ must contain this term as well, because $V_{max}$ is a term of maximal interaction size in $V$ and therefore its contribution to the commutator $i[X,V]$ cannot be canceled by other terms in $V$.  But for $k_{max} > 1$, we can always choose $k' \leq k$ such that $X$ exists and $k_{max} + k' - 1 > k$. Then $i[X,V]$ is not $k$-local and we have a contradiction. It follows that $k_{max} = 1$, i.e. $V$ is a sum of single-site operators.

The global version of this lemma easily follows.  That is, we can now argue that any unitary operator $U$ that preserves the subspace of $k$-local Hamiltonians under conjugation must be generated by local (single-site) unitary operators and permutations of sites.  Assume $U$ preserves the subspace of $k$-local Hamiltonians.  Consider any 1-local operator $O$. Then for any $k$-local operator $X$, $U^+XU$ is $k$-local, so $i[O,U^+XU]$ is $k$-local also.  Because $U$ preserves the subspace of $k$-local operators, $iU[O,U^+XU]U^+=i[UOU^+,X]$ is $k$-local as well.  That is, $C_{UOU^+}$ preserves the subspace of $k$-local Hamiltonians.  By the ``linearized" version of the lemma above, $UOU^+$ must then be 1-local.  That is, $U$ must send 1-local operators to 1-local operators under conjugation.  Thus for any single-site operator $O$, $UOU^+$ must be a sum of single-site terms.  By considering $(UOU^+)^2$, which must also be a sum of single-site terms, we see that $UOU^+$ must have support on a single site.  Moreover, for two single-site operators $O$ and $O'$ on the same site, $UOU^+$ and $UO'U^+$ must also be on the same site, otherwise $U(OO')U^+$ would not be 1-local.  We conclude that conjugation by $U$ sends the algebra of operators on one site to the algebra of operators on another site; it follows that $U$ is generated by a combination of single-site unitary operators and permutations of sites.

\subsubsection{Finite number of duals}

As discussed above, we would like to show that generic $k$-local Hamiltonians do not have $k$-local duals.  However, the statements proven in this paper will be weaker statements.  More specifically, our result will apply to any particular linear subspace of Hamiltonians, for a Hilbert space of fixed size.  For instance, consider the subspace of all $k$-local Hamiltonians on $n$ qubits.  The result then states: if there exists a single example of a $k$-local Hamiltonian on $n$ qubits without any duals, then almost all $k$-local Hamiltonians on $n$ qubits do not have $k$-local duals.  The analogous result applies to systems of qudits rather than qubits (i.e. using $d$-dimensional subsystems). Or, for instance, consider the linear subspace of all Hamiltonians on spin chains with nearest-neighbor couplings.  Then the result states: if there exists a single example of a spin chain Hamiltonian on $n$ spins without any duals that are also spin chains, then the same must hold for almost all spin chain Hamiltonians on $n$ spins.  

Of course, these results on their own do not guarantee that almost all $k$-local Hamiltonians do not have duals.  However, as a proof of principle, we will numerically find an example of a translation-invariant spin chain Hamiltonian on 10 spins that does not have any translation-invariant spin chain duals.  Combined with the above result, the numerical example effectively proves that almost all translation-invariant spin chain Hamiltonians on 10 spins do not have any translation-invariant spin chain duals. 

If a family of examples could be generated analytically for systems of different sizes, perhaps using induction in the size of the system, then our result would imply rigorously that almost all $k$-local Hamiltonians do not have duals.  We suspect that such examples exist, which would imply that the general result holds.

Now we begin to formalize the statement.  Consider a subspace of Hamiltonians, $S \subset \Herm(\HH)$.  For instance, $S$ may be the subspace of $k$-local Hamiltonians on $n$ qubits.  For a given local Hamiltonian $H \in S$, we are interested in whether $H$ has a dual $H' \in S$.  By dual, we mean a Hamiltonian $H'$ with the same spectrum as $H$, such that $H'$ is not related by any combination of local unitary transformations, permutations of qubits, or the transpose operation.  Let $G$ be the subgroup of linear transformations on $S$ generated by local unitary operations, permutations of qubits, and the transpose operation.  In addition, let the unitary group $U(N)$ act on $\Herm(\HH)$ by conjugation.  Note that the orbit $\Orb_{U(N)}(H)$ is the set of Hamiltonians isospectral to $H$. Then the local duals of $H$ are precisely the points in $\Orb_{U(N)}(H) \cap S$ that are not in $\Orb_{G}(H)$.  Note that $ \Orb_{G}(H) \subset S \cap \Orb_{U(N)}(H).$  The statement that $H$ has no duals is the statement that 
\begin{equation}
S \cap \Orb_{U(N)}(H) = \Orb_{G}(H)\,.
\end{equation}
This condition says that the only Hamiltonians in $S$ isospectral to $H$ are those related by local unitary operations, permutations of qubits, and the transpose operation.  Equivalently, the condition states $H \in S$ is uniquely determined by its spectrum, up to the previous operations.  The situation of a Hamiltonian with no duals is illustrated in Figure 1.
\begin{figure}[t] \label{fig:wacky}
  \centering
    \includegraphics[width=.6\textwidth]{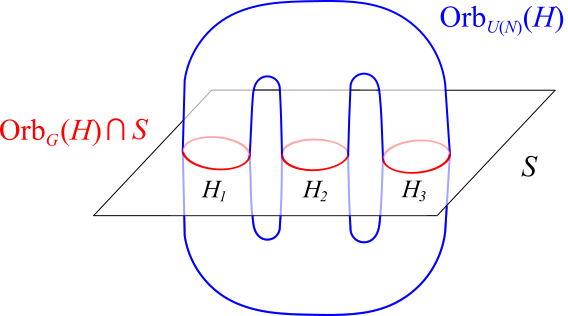}
     \caption{We depict the spaces $S$ and $\Orb_{U(N)}(H)$ intersecting in the ambient space $\Herm(\HH)$. The Hamiltonian $H$ is depicted to have no duals.  The orbit intersects $S$ in multiple disconnected components, appearing as circles in the diagram.  These disconnected components together make up $ \Orb_{G}(H)  =S \cap \Orb_{U(N)}(H).$  Each disconnected component of $S \cap \Orb_{U(N)}(H)$ contains Hamiltonians related by local unitaries (which are continuous transformations), and the sets are related to one another by permutations of qubits and transposition, which are discrete transformations.  Alternatively, if the intersection contained points not related to $H$ by local unitaries, permutations, or transposition, then $H$ would have duals. The figure is only intended as a schematic representation of the spaces involved.}
\end{figure}

As a first step, we can constrain the answer by counting the dimensions of the spaces involved.  How big is $\dim(\Orb_{U(N)}(H))$?  For simplicity, consider the tangent space of $\Orb_{U(N)}(H)$ at $H$, given by the image of the linear map $C_H : \Herm(\HH) \to \Herm(\HH)$, taking $V \mapsto i[V,H]$.  The kernel of $C_H$ will be the Hermitian operators that commute with $H$, i.e.\! that are diagonal in the energy eigenbasis.  A generic local Hamiltonian will have non-degenerate spectrum, so there will be an $N$-dimensional space of these operators.  Hence $C_H$ has rank $N^2-N$ (using $\dim \Herm(\HH) = N^2$), and likewise $\Orb_{U(N)}(H) = N^2-N$.  Meanwhile, $\dim S \ll N$, so $\dim S + \dim \Orb_{U(N)}(H) \ll \dim \Herm(\HH)$.  This inequality makes it possible for $S \cap \Orb_{U(N)}(H)$ to be empty, because the ambient space $ \Herm(\HH)$ is sufficiently large.  We already know $\dim S \cap \dim \Orb_{U(N)}(H) \supset \Orb_{G}(H)$, so the intersection is not empty, but we might still expect $\dim S \cap \dim \Orb_{U(N)}(H)$ contains no further points.

Concretely, we will consider perturbing a Hamiltonian $H$ in the subspace $S$ by unitary conjugation, such that the perturbed Hamiltonian still lies within $S$.  We wish to show that for almost all $H$, such unitary perturbations will either commute with the Hamiltonian (and so have no effect at all), or otherwise be local unitary perturbations.  This generically precludes arbitrarily nearby duals in the sense that for almost all $H$ in $S$, there is an open set in $S$ containing $H$ for which there are no duals of $H$.

To implement the proof, we construct a complex-valued polynomial function of the Hamiltonian, $p(H)$, which has roots precisely at the values of $H$ that have infinitesimally nearby duals in $S$.  However, any polynomial has the property that either its zeros are a set of measure zero, or the polynomial is identically zero.  In our context, this means that either \textit{almost all} Hamiltonians do not have arbitrarily nearby duals, or \textit{all} Hamiltonians have arbitrarily nearby duals.  Thus, we only need to check if one Hamiltonian has no arbitrarily nearby duals to determine which scenario holds for $S$.

\begin{figure}[t]
  \centering
    \includegraphics[width=0.65\textwidth]{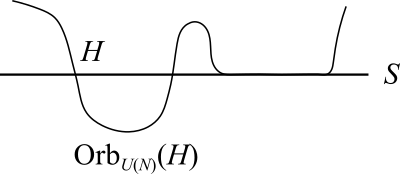}
     \caption{We visualize what \textit{cannot} happen, namely the intersection of $\Orb_{G}(H)$ and $S$ contains both isolated points \textit{and} open sets.  The directions of $S$ which correspond to local unitary perturbations of the Hamiltonian are suppressed.}
\end{figure}

Figure 2 depicts a scenario we have shown impossible, namely a Hamiltonian in $S$ which is dual to a proper open subset of $S$.  Our theorem is in fact stronger: if a single Hamiltonian in $S$ does not have arbitrarily nearby duals, then almost all other Hamiltonians in $S$ have no arbitrarily nearby duals.  Indeed, the weaker statement follows since if there were an open set of duals, the number of Hamiltonians with no arbitrarily nearby duals would \textit{not} be measure zero.

We now proceed with the result:

\begin{thrm} [Finite number of duals] \label{thrm:finiteduals}
Suppose that we have a subspace $S$ of Hermitian matrices $\Herm(N)$  together with some subgroup $G \subset U(N)$ that preserves $S$ when acting by conjugation. Moreover suppose that there exists a matrix $H_0 \in S$ whose Jordan form is the generic Jordan form\footnote{By generic Jordan form on $S$, we mean the unique Jordan normal form associated to all operators in a non-empty, Zariski open subset of $S$.  For most subspaces $S$ of local Hamiltonians we are interested in, the generic Jordan form will be non-degenerate, but we consider the general case here.  For instance, translation-invariant spin chains with periodic boundary conditions are generically degenerate.} on $S$, such that if
$$i[V,H_0] \in S$$
for some Hermitian matrix $V$ then either $[V,H_0] = 0$ or $V \in \mathfrak{g}$, where $\mathfrak{g}$ is the Lie algebra\footnote{We use the physicist's convention of taking the Lie algebra to consist of Hermitian rather than anti-Hermitian operators.} of $G$.  (In particular, $\mathfrak{g}$ is equal to the space of 1-local Hamiltonians.) 

\noindent Then the same property holds for almost all matrices $H \in S$: if
$$i[V,H] \in S$$
for some Hermitian matrix $V$, then either $[V,H] = 0$ or $V \in \mathfrak{g}$.\footnote{The above also result also holds if one replaces $S \subset \Herm(N)$ with the complexification \\$S_\C \subset \Mat(N,\C)$ and replaces $G \subset U(N)$ with the complexification $G_\C \subset GL(N)$.  The complexified version of the statement holds by a nearly identical argument, and the complexified statement will be used in the proof of Theorem \ref{thrm:constduals}.}

\noindent In particular, if $S$ is some space of local Hamiltonians with $G$ the group of local unitary operators, then almost all Hamiltonians $H \in S$ have a finite number of duals.

\end{thrm}

\noindent \textit{Proof}. Consider the linear map $f_H : \Herm(\HH) \to S^\perp$ on the space of Hermitian matrices, defined by
\begin{align}
f_H(V) = \mathrm{Proj}_{\,S^\perp}\, i[V,H]
\end{align}
where the projector is onto the orthogonal complement of $S$ with respect to the Hilbert-Schmidt inner product. Then $\ker(f_H)$ is the space of Hermitian matrices $V$ such that
\begin{align}
i[V,H] \in S\,.
\end{align}
By construction, $\mathfrak{g}\subset \ker(f_H)$ and $\Stab_{\mathfrak{u}}(H) \subset \ker(f_H)$, where  \begin{align}
\Stab_{\mathfrak{u}}(H) = \{V \in \mathfrak{u}=\Herm(\HH) \, | \, [V,H]=0 \}
\end{align}
is the stabilizer of $H$ under the adjoint action of the Lie algebra $\mathfrak{u}$ of $U(N)$.  So for any $H$, we have
\begin{align}
\mathfrak{g}  \cup \Stab_{\mathfrak{u}}(H) \subset &  \ker(f_H) \\
\dim \left(\mathfrak{g}  \cup  \Stab_{\mathfrak{u}}(H) \right) \leq & \dim \ker(f_H) \nonumber \,.
\end{align}
Also note that
\begin{align}
\dim \left(\mathfrak{g}  \cup  \Stab_{\mathfrak{u}}(H) \right)=\dim \mathfrak{g} + \dim  \Stab_{\mathfrak{u}}(H) - \dim \left(\mathfrak{g} \cap \Stab_\mathfrak{u}(H)  \right)
\end{align}
and $\mathfrak{g} \cap \Stab_\mathfrak{u}(H)  =  \Stab_\mathfrak{g}(H)$, so 
\begin{align}\label{eq:loc1}
\dim \ker(f_H) \geq \dim \left(\mathfrak{g} \cup \Stab_{\mathfrak{u}}(H)\right) = \mathrm{dim}\, \mathfrak{g} +\dim \Stab_{\mathfrak{u}}(H) - \dim \Stab_\mathfrak{g}(H) \, .
\end{align}

We now argue that for almost all  $H \in S$, $\dim \ker(f_H)$ attains its minimal value over $S$, i.e.
\begin{align}
\dim \ker f_H = \min_{H' \in S} \dim \ker f_{H'} 
\end{align}
for almost all $H \in S$.  This fact immediately follows from the lower semicontinuity of the rank of a matrix, but we will elaborate to provide some intuition about the role of polynomials.  First, recall that a matrix $M$ has rank at least $r$ if and only if there exists an $r \times r$ matrix minor of $M$ with nonzero determinant.  Now suppose that $\dim \ker f_{H_0} = r$ for some $H_0$. If we consider $f_{H_0}$ as a matrix, there must exist some matrix minor of $f_{H_0}$ with nonzero determinant formed by removing $r$ rows and columns from $f_{H_0}$.  The determinant will be a polynomial function of the entries of $H_0$, and because a non-zero polynomial is non-zero almost everywhere, the determinant will be non-zero almost everywhere in $S$.  So for any $H_0 \in S$, and for almost all $H \in S$, 
\begin{align} \label{eq:loc2}
\dim \ker H \leq \dim \ker H_0 \, .
\end{align}

Now let us assume there exists an $H_0$ satisfying the hypotheses of the theorem, which may be rewritten as $\ker f_{H_0} = \mathfrak{g} \cup \Stab_{U(N)}(H_0)$, along with the requirement that $H_0$ has the generic Jordan form on $S$.  Combining Equation \ref{eq:loc2} with Equation \ref{eq:loc1}, we then have that for almost all $H \in S$,
\begin{align} \label{eq:loc5}
\dim \ker f_H \leq \dim \ker  f_{H_0} = \mathrm{dim}\, \mathfrak{g} +\dim \Stab_{\mathfrak{u}}(H_0) - \dim \Stab_\mathfrak{g}(H_0).
\end{align}
Because $H_0$ is assumed to have the generic Jordan form on $S$, and because the generic Jordan form will have the smallest eigenvalue degeneracy, $H_0$ will then have the smallest stabilizer under adjoint action, i.e.
\begin{align} \label{eq:loc3}
\dim \Stab_\mathfrak{u}(H_0) = \min_H \dim \Stab_\mathfrak{u}(H).
\end{align}
Meanwhile, $\dim \Stab_\mathfrak{g}(H) = \dim \ker g_H$ where $g_H(V)= [V,H]$, so by the exact same argument made below Equation \ref{eq:loc2}, 
\begin{align}  \label{eq:loc4}
\dim  \Stab_\mathfrak{g}(H) = \min_{H' \in S}  \Stab_\mathfrak{g}(H')
\end{align}
for almost all $H \in S$.

Plugging Equations \ref{eq:loc3} and \ref{eq:loc4} into Equations \ref{eq:loc5} and \ref{eq:loc1}, one finds that for almost all $H \in S$,
\begin{align}
\dim \ker f_H \leq &  \dim \mathfrak{g} + \min_{H'} \dim \Stab_{\mathfrak{u}}(H') -  \min_{H'} \dim \Stab_\mathfrak{g}(H') \\
\dim \ker f_H \geq &\dim \mathfrak{g} + \min_{H'} \dim \Stab_{\mathfrak{u}}(H') -  \min_{H'} \dim \Stab_\mathfrak{g}(H') \nonumber
\end{align}
and hence the above is an equality, with 
\begin{align}
\dim \ker f_H = \dim (\mathfrak{g} \cup \Stab_\mathfrak{u}(H))
\end{align}
which implies
\begin{align}
\ker f_H  = \mathfrak{g} \cup \Stab_\mathfrak{u}(H)
\end{align}
because $\ker f_H \supset   \mathfrak{g} \cup \Stab_\mathfrak{u}(H)$.  Finally, the above expression is precisely the desired condition of the theorem.

To complete the proof of Theorem \ref{thrm:finiteduals}, it only remains to show that the number of duals is finite. Suppose that for some $H \in S$ then for any $V \in \Herm(N)$, $i[V,H] \in S$ implies either that $[V,H] = 0$ or $V \in \mathfrak{g}$. Then there must exist a finite volume around the identity in $U(N)$ within which $UHU^\dagger \in S$ implies $U \in G$ or $UHU^\dagger = H$. However since the unitary group is compact, this can only be true for every $H' = UHU^\dagger \in S$ if the set of Hamiltonians in $S$ with the same eigenvalues as $H$ quotiented by the action of $G$ by conjugation is finite. However we have already shown that this exact result is true for almost all $H \in S$. It follows that for almost all $H \in S$, the set of Hamiltonians in $S$ with the same eigenvalues as $H$ quotiented by the action of $G$  is finite, which we defined to be the number of duals.

\subsubsection{Constant number of duals} \label{sec:constduals}

To extend our result to non-infinitesimal unitary transformations, considering the whole orbit $\Orb_{U(N)}(H)$ rather than just the tangent space at $H$, we make use of more sophisticated mathematical tools than were necessary for the previous results. The proof consists mostly of classical algebraic geometry, though it makes use of some theorems phrased in the language of schemes.\footnote{The maps and spaces involved are also complex analytic, a weaker condition.  While analyticity alone should be sufficiently strong to prove the result here, we make use of algebraic structure instead.} Nevertheless the basic strategy, as well as the result itself, is highly analogous to the previous section. We show that almost all local Hamiltonians have the same number of duals.  That is, the number of duals per Hamiltonian is almost everywhere constant over the space of local Hamiltonians.  The numerical results in Section \ref{sec:numerics} will augment the theorem below to show that the number of duals is generically zero (rather than simply being constant), at least for certain small systems.

One main difference from the style of the previous proof is that we must include non-Hermitian local Hamiltonians when searching for duals, rather than just ordinary Hermitian Hamiltonians. In other words, we must consider the orbit of $H$ under conjugation with $GL(N)$ and not just $U(N)$. Similarly, we generalize the equivalence classes associated to a single TPS to include the orbit under conjugation by elements of $GL(d)$ on each subsystem (analogous to local unitaries), as well as the familiar permutation of the subsystems and transposition. This requirement is particularly important as it means we are working over an algebraically-closed field, the complex numbers. When performing associated numerics, the complexification adds a small amount of numerical difficulty, since we must search a space with twice the number of parameters.

Although the proof itself is somewhat technical, the outline is easy to understand. First we construct the space of orbits of local complex Hamiltonians under conjugation by local operators on each subsystem, permutation of subsystems, and transposition.  Then we define a map from this space such that two orbits are mapped to the same point if and only if they have the same spectrum and are therefore dual.  Here, the complexification of the spaces becomes important. Note that the number of distinct solutions to the complex algebraic equation $f(z) = k$ is the same for almost all values of $k$, although the analogous statement does not hold for real solutions.  Similarly, for a class of sufficiently well-behaved maps, the number of points in the fiber will be constant almost everywhere.\footnote{We first need to projectivize the spaces in order for this statement to hold.}

\begin{thrm}[Constant number of duals] \label{thrm:constduals}
Suppose that we have a complex subspace $S_{\C}$ of matrices $\Mat(N,\C)$ that is preserved by transposition, together with some reductive subgroup $G_0 \subset GL(N)$ that preserves $S_{\C}$ when acting by conjugation and is invariant under transposition. Let $G$ be the subgroup of $GL(N^2)$ whose fundamental representation is generated by transposition and the action of $G_0$ by conjugation.  Suppose that almost all matrices in $S_{\C}$ are diagonalizable and moreover that for almost all matrices in $S_{\C}$, the number of $G$-orbits $S_{\C}$ which are similar is finite; we refer each such orbit as a  ``complex dual." Then the number of complex duals is constant on a Zariski open subset of $S_{\C}$.
\end{thrm}

\noindent \textit{Proof}. We want to define a morphism of varieties for which the domain is the orbits of $S_{\C}$ under $G$ and for which the fibers are the sets of duals.  Our starting point is the rational map $f: \mathrm{Proj} \,(S_{\C}) \to \Lambda$, defined to be the projectivization of the map
$$H \longmapsto \{\mathrm{Tr}(H^n) \, | \, 1 \leq n \leq N\}\,.$$
Here $\Lambda$ can be taken to be the weighted projective space which  is the quotient of  $\C^N \setminus \{0\}$  by the action of the multiplicative group $\C^*$ of nonzero complex numbers, taking
$$ (t_1,t_2,...,t_N) \mapsto (\lambda t_1, \lambda^2 t_2, ... ,\lambda^N t_N)\,.$$
Note that we can identify the quotient of $\C^N$ by the permutation group $S_N$ (acting by permuting indices) with $\C^N$ itself via the map
$$ (\lambda_1, \lambda_2, ... ,\lambda_N) \mapsto  (e_1,e_2,...,e_N)$$
where $e_j(\lambda_1,...,\lambda_N)$ are the symmetric polynomials of $\{\lambda_i\}$, which may be defined by matching coefficients of the formal power series
$$ \prod_{j=1}^N (x + \lambda_j) = x^N + \sum_{j=1}^N e_{N-j}x^j.$$
The elementary symmetric polynomials $\{e_j\}$ of $\{\lambda_j\}$ can then be identified via Newton's identities with the power sum symmetric polynomials $\{t_j\}$. Since $t_n = \mathrm{Tr}(H^n)$ is the $n$th power sum symmetric polynomial of the eigenvalues $\{\lambda_j\}$ of $H$, this gives us an identification of   $f: \mathrm{Proj} \,(S_{\C}) \to \Lambda$ with the map that associates to a matrix $H$ its  projectivized set of eigenvalues with algebraic (not geometric) multiplicities for $H$. We make use of projective rather than affine spaces in this construction simply because we later need to take advantage of the nicer properties of projective morphisms. We are assuming that $S_{\C}$ is nonzero and the generic Jordan normal form of a matrix in $S_{\C}$ is diagonalizable, so a dense open subset in  the projective variety $\overline{\mathrm{Im}(f)}$, which is the closure in $\Lambda$ of the image of $f$, has fibers that are the intersection of an orbit of $GL(N)$ with $S_{\C}$, a set of similar matrices.

Now we want to quotient $ \mathrm{Proj} \,(S_{\C})$ by the action of $G$. Since $G$ is not compact, the topological quotient of $\mathrm{Proj} \,(S_{\C})$ by $G$ is not well behaved. Instead we will show that a GIT (geometric invariant theory) quotient exists. A GIT quotient of a projective variety is well-defined for a linearized action of a reductive algebraic group. Since $G_0$ was assumed to be reductive and $G$ is a finite extension of $G_0$ which acts linearly on $S_{\C}$ we have a linearized action of $G$ on $\mathrm{Proj} \,(S_{\C})$ and  we can construct a GIT quotient by $G$. A  GIT quotient of a projective variety $X$ with a linear action of $G$ gives a categorical quotient $X^{ss} \to X^{ss}\,/\!\!/\,G$ where $X^{ss}$ are the semistable points of $X$. A point  $H$ is semistable  if and only if there exists a homogeneous $G$-invariant polynomial which is non-zero at $H$. In our case, for any $H$ that is not nilpotent, $\Tr (H^n)$ is $GL(N)$-equivariant and homogeneous and will be non-zero for some $n$.

Since a GIT quotient is a categorical quotient on the category of algebraic varieties, any $G$-invariant map will uniquely factor through the quotient. Since, for all $n$, $\mathrm{Tr}(H^n)$ is $G$-invariant, the restriction of $f$ to the semistable points will uniquely factor through a map we shall refer to as $f_2: \mathrm{Proj} \,(S_{\C})^{ss} \,/\!\!/\, G \to\Lambda$. A GIT quotient is only a geometric quotient on an open subset known as the stable points, here used in the original Mumford sense \cite{GIT}[Definition 1.7]. However since the number of $G$-orbits mapped to a given point in $\Lambda$ is generically finite, such generic orbits will be stable, because all the orbits in a $G$-invariant open neighborhood given by the inverse image under $f$ of an open subset of $\Lambda$ will be closed, as the fibers of $f$ will all be closed and the fibers are finite disjoint unions of orbits. It follows that the open subset of stable points is non-empty and hence dense (since $\mathrm{Proj} \,(S_{\C})^{ss} \,/\!\!/\, G$ is irreducible).

The next step will be to take the projective morphism $f_2$ and base change onto an open subscheme of the image of $f_2$. This new morphism will still be projective since projectiveness is preserved under base changes.

We make use of  \cite{Hartshorne} III Corollary 10.7 which states that if $f:X \to Y$ is a morphism of non-singular varieties over an algebraically closed field $k$ of characteristic $0$, then there is a non-empty open subset $V \subseteq Y$ such that $f:f^{-1}(V) \to V$ is smooth. This corollary does not directly apply to $f_2$ since $X = S_\C \,/\!\!/\, G$ is not necessarily nonsingular, but since we have assumed that almost all the fibers are finite, the dimension of the image must be equal to the dimension of $X$. This means that the image of the singular locus $\text{Sing}(X)$ is not dense in the image of $X$ and hence the complement of the closure of $f_2(\text{Sing}(X))$ in the closure of $f_2(X)$ will be an non-empty open subset $V_1 \subseteq Y$ and have non-singular preimage in $X$. Then applying Corollary 10.7 to the restriction of $f_2$ to $f_2^{-1}(V_1)$, we learn that there exists an non-empty open subset $V_2 \subseteq X$ such that the restriction of $f_2$ to $V_2$ is smooth.

If a morphism is smooth, it is also flat (\cite{Hartshorne} III Theorem 10.2). Further, $f_2(V_2)$ is a Noetherian scheme since it is quasi-projective. This mean that the degree of the fiber, which for finite fibers is just the number of points in the fiber (counting multiplicities) is constant everywhere (\cite{Hartshorne} III Corollary 9.10).

Now we take the intersection of $V_2$ with the stable points $S_\C \,/\!\!/\, G$. The restriction of the morphism $f_2$ to this dense open subset will be a morphism from the geometric quotient by $G$ and hence the degree of the fibers will simply count the number of geometric orbits. We have therefore shown that a dense open subset of local complex Hamiltonians  have a constant number of complex duals, which completes the proof of Theorem \ref{thrm:constduals}.

Notice that  the intersection of any nonempty Zariski open with the real subspace of a complex vector space has complement of measure zero in the real vector space, since the Zariski open is  the complement of the solution space of a set of complex algebraic equations. When we restrict to the real subspace this becomes the complement of a set of real algebraic equations (the real and imaginary parts of the original equations) and all real algebraic equations have measure zero solution except for $0=0$. If there exists a Zariski open of local complex Hamiltonians with no complex duals, then generic local (real) Hamiltonians have no complex duals, and hence since Hermitian duals are simply a subclass of complex duals, they also have no Hermitian duals.

We also need to show that the assumptions that we made for $S_{\C}$ and $G_0$ apply for the particular case of a space of local complex Hamiltonians with conjugations by local $GL(d)$ and permutations of tensor product factors.

$S_{\C}$ and $G_0$ are trivially invariant under transposition. To show that $S_{\C}$ is generically diagonalizable we first note that in any Zariski closed subspace of $\Mat(N,\C)$, matrices will have the generic Jordan form for that space on a Zariski open subspace of it. Then exactly the same arguments as above, tell us that, generically, Hermitian matrices in $S_{\C}$ will have the generic Jordan form for $S_{\C}$.  Since all Hermitian matrices are diagonalizable, the generic Jordan form for $S_{\C}$ must be diagonalizable. To show that $G_0$ is reductive, we note that the connected component of $G_0$ is the direct product of copies of $GL(d)$ quotient by a subgroup of the center of the direct product group. It then follows that since $GL(d)$ is reductive, so is $G_0$.

Finally we need to be able to confirm that any example we might construct (either numerically or analytically) with no complex duals, lies in the open subset of local Hamiltonians in which the number of duals is constant. Firstly, we note from our proof of Theorem \ref{thrm:constduals} that if $H$ satisfies the conditions in the complex version of Theorem \ref{thrm:finiteduals} then it lies in the open subspace of stable points of $S_{\C}$. We then simply need to show that $f_2$ is smooth at $H$. However since the GIT quotient is locally just a geometric quotient, this will be true so long as the differential of $f$ is surjective on the tangent space of $S_{\C}$ quotiented by the tangent space of the orbit of $G_0$, which is again just a restatement of the conditions for Theorem \ref{thrm:finiteduals}. Finally, again because the differential is surjective, the point in the fiber necessarily has trivial multiplicity.

Combining Theorems \ref{thrm:finiteduals} and \ref{thrm:constduals}, we have therefore proved analytically that, if we have a single example (subject to the conditions described above) in some class of local Hamiltonians which has no complex duals, then almost all Hamiltonians in that class have a unique TPS in which they are local.

\section{Numerics} \label{sec:numerics}

In the previous section we showed that for systems of a fixed size, if you can find a single example of a local Hamiltonian with a unique local TPS, then generic local Hamiltonians of that size must also have a unique local TPS.  In this section, we use numerics to demonstrate that such ``example Hamiltonians" exist, at least for a small class of numerically tractable problems.  These numerical examples, when combined with Theorems \ref{thrm:finiteduals} and \ref{thrm:constduals}, amount to a proof of the following statements:

\begin{enumerate}
\item  Almost all 2-local Hamiltonians on 10 qubits have finitely many (and possibly zero) 2-local duals.
\item  Almost all nearest-neighbor Hamiltonians on 10-qubit spin chains have finitely many (and possibly zero) 2-local duals.
\end{enumerate}
The above statements are fully proven, if the associated numerical calculation is robust.  We believe the numerical result that aids the proof (analogous to numerically calculating that a certain quantity is nonzero) is robust to finite-precision machine error, although we do not undertake a rigorous analysis of the error.   On the other hand, the result below is only verified in a probabilistic fashion, as elaborated later in the section.
\begin{enumerate}
\setcounter{enumi}{2}
\item (\textit{Probabilistically verified}) Almost all translation-invariant, nearest-neighbor Hamiltonians on 6-qubit spin chains have no translation-invariant duals.
\end{enumerate}
The numerical calculations behind these results are discussed below.

\subsection{Example showing finite duals}
First we focus on finding an example of a local Hamiltonian with a finite number of duals, or equivalently, a local Hamiltonian without infinitesimally nearby duals.  That is, we want a Hamiltonian that will satisfy the hypotheses of Theorem \ref{thrm:finiteduals}.  The theorem applies within the context of a fixed Hilbert space $\HH$ and a fixed subspace $S$ of local Hamiltonians, such as the subspace of 2-local Hamiltonians on 10 qubits ($k$=2, $d$=2, $n$=10).  A valid ``example Hamiltonian" $H_0$ must have non-degenerate spectrum, and it must have the property listed in Theorem \ref{thrm:finiteduals}: for any $V \in \Herm(\HH)$ such that $i[V,H_0] \in S$, either $[V,H_0]=0$ or $V \in \mathfrak{g}$.  Given an example Hamiltonian, application of Theorem \ref{thrm:finiteduals} implies Statement 1 above.

We want to choose a particular Hamiltonian $H_0 \in S$ and check numerically that the above criterion  holds.  From the proof of Theorem \ref{thrm:finiteduals}, we see that the criterion is equivalent to asking that $\dim(\ker f_H) = \dim G - \dim G_C(H_0) + N$, provided that $H_0$ has nondegenerate spectrum.  The connected component of $G$ is the group of local unitary operators, so $\dim G = n(d^2-1)+1$.  We will assume that $H_0$ does not commute with any local unitaries besides the identity, which is true for generic local Hamiltonians, and which is easy to check for a particular $H_0$.  Then $\dim G_C(H_0)=1$, and the criterion becomes
\begin{equation}
\dim(\ker f_{H_0}) =n(d^2-1) +  N.
\end{equation}

For a particular choice of $H_0$, one could compute the rank of the operator $f_{H_0}$ directly.  However, we will use a more efficient approach to check the above criterion.  Note that 
\begin{equation}
C_H(\ker f_{H_0}) = \Imm(C_{H_0}) \cap S
\end{equation} and 
\begin{equation}
\dim  \Imm(C_{H_0}) \cap S = \dim C_H(\ker f_{H_0})  =\dim (\ker f_{H_0}) -N.
\end{equation} Furthermore, $\Imm(C_H) = \{[A,H] \, | \, A \in \Herm(\HH)\}$ is precisely the set of operators with zero diagonal entries in the eigenbasis $\{\ket{E_i}\}$ of $H_0$.  That is, 
\begin{equation}
\Imm(C_H)  = \{A \,|\, A \in \Herm(\HH) \text{ s.t. } \bra{E_i}A\ket{E_i} \text{ for } i=1,..,N\}.
\end{equation} So $ \Imm(C_{H_0}) \cap S$ is the set of local operators that have all zero diagonal entries in the eigenbasis of $H_0$.  With this motivation, define the matrix
\begin{equation}
M_{ij}:=\bra{E_i} L_j \ket{E_i}
\end{equation} where $\ket{E_i}$ are the eigenvectors of $H_0$, and $\{L_i\}_{i=1}^s$ in some basis for $S$.  The matrix $M$ may be computed somewhat efficiently.  The vectors in $\ker M$ correspond to elements of $\Imm(C_{H_0}) \cap S$.  (Alternatively, it is easy to see that $\ker M$ is the set of local operators which, when added to the Hamiltonian, do not alter the spectrum to first order in perturbation theory.)  Thus $\dim \ker M = \dim \ker f_{H_0}$, and checking the  criterion of Theorem \ref{thrm:finiteduals} for $H_0$ only requires computing $\dim \ker M $, with the criterion satisfied if $\dim \ker M =n(d^2-1)$.
	
Although the chosen $H_0$ need not be ``random," we may choose $H_0$ by randomly generating a 2-local Hamiltonian on 10 qubits ($k$=2, $d$=2, $n$=10).  First we must confirm numerically that the spectrum is nondegenerate.  Then we must confirm that $H_0$ does not commute with any local unitary operators besides the identity.  (This condition can be checked analytically for any particular $H_0$\,.)  Finally, we must calculate $ \ker M$.  

Numerically, we calculated $\dim \ker M =n(d^2-1) +  N=30.$  Then applying Theorem \ref{thrm:finiteduals}, we have effectively proven that almost all 2-local, 10 qubit Hamiltonians have a finite number of 2-local duals (and possibly no duals).

Again, the above conclusion only applies to \textit{almost all} 2-local  Hamiltonians, 10-qubit Hamiltonians.  Any \textit{particular} 2-local Hamiltonian may well have an infinitude of duals, but such Hamiltonians are measure zero.  However, we can still make conclusions about more specific classes of Hamiltonians. Consider some more narrow subclass of local Hamiltonians, given as a linear subspace $V \subset S \subset \Herm(\HH)$.  For instance, we might be interested in taking $V$ as the subspace of Hamiltonians with nearest-neighbor couplings on spin chains, while $S$ is still the space of 2-local Hamiltonians.  Because $V$ is measure zero within $S$, the above result still allows the possibility that all Hamiltonians in $V$ have infinitely many 2-local duals.   However, a simple extension  of Theorem \ref{thrm:finiteduals} implies that given an example Hamiltonian $H_0 \in V \subset S$, we can conclude not only that almost all Hamiltonians in $S$ have finitely many duals, but also that almost all Hamiltonians in $V$ have finitely many duals, which is a stronger statement. 

We randomly generated examples of Hamiltonians with nearest-neighbor couplings on 10-qubit spin chains.  These examples satisfied the hypothesis of Theorem \ref{thrm:finiteduals}, taking $S$ to be the space of 2-local Hamiltonians on 10 qubits.  Then the above generalization of Theorem \ref{thrm:finiteduals} implies that almost all 10-qubit spin chain Hamiltonians have finitely many 2-local duals.  

\subsection{Example showing no duals}
Now we find an example to satisfy the hypotheses of of Theorem \ref{thrm:constduals}, in order to verify Statement 3 above.  That is, we want an example of Hamiltonian $H_0 \in S$ such that $H_0$ has no duals in $S_\C$, where $S_\C$ is the complexification of $S$. Because this task is more difficult numerically, we choose a smaller subspace for $S$.  In particular, we will consider $S$ as the space of translation-invariant, nearest-neighbor Hamiltonians on 6-qubit spin chains; $S_\C$ consists of complex linear combinations of these operators.  Although the non-Hermitian operators in $S_\C$ do not correspond to physical Hamiltonians, we are interested in checking for complex duals in order to satisfy the hypotheses of Theorem \ref{thrm:constduals}, which then applies to generic  local Hamiltonians that \textit{are} Hermitian.

Given a randomly generated Hamiltonian in $H_0 \in S$, we want to find all possible complex duals by searching over the space $S_\C$ for (non-Hermitian) Hamiltonians with the same spectrum.  That is, we want to find solutions $H$ to the equation 
\begin{equation}
\textrm{spectrum}(H) = \textrm{spectrum}(H_0)
\end{equation}
for $H \in S_\C$, where spectrum($H$) denotes the list of eigenvalues.  These equations are difficult to solve, and the problem falls in the general class of ``inverse eigenvalue problems" \cite{Chen1}.  It is useful to think of the solutions as solving an optimization problem: the duals are given by the set
\begin{equation}
\underset{H \in S_\C}{\operatorname{arg min}}\, \| \textrm{spectrum}(H)-\textrm{spectrum}(H_0) \|
\end{equation}
up to equivalence by local unitary operators and translations and reflections of the chain.  Any norm may be used (for instance, the $\ell_2$--norm).  The duals will be exact global minima, with $\| \textrm{spectrum}(H)-\textrm{spectrum}(H_0)\|=0$.  

Because we only care about finding duals up to conjugation by 1-local operators, it would be more efficient to search over a quotient of $S_\C$ by the action of 1-local operators, rather than searching over the full space $S_\C$.  In fact, such a quotient is easy to define.  For simplicity, we will first describe how to form a quotient of $S$ by local unitary operators, rather than a quotient of $S_\C$ by general 1-local operators.  Moreover, we will consider the case of qubits in a translation-invariant spin chain, but the construction is easily generalized to qudits with any pattern of interactions.

By analogy to the language of gauge theory and gauge-fixing, one might say that we want to find a``gauge-fixed" $W \subset S$, where local unitary operators on qubits play the role of the gauge group. This analogy may be helpful to those familiar; otherwise, we simply say that $W$ should have exactly one representative from each orbit of the group of local unitaries acting on $S$.  We can write $H \in S$ as 
\begin{equation}\label{eq:spinchain} 
H=\sum_{i=1}^n \sum_{a=0}^3 \sum_{b=1}^3 c_{ab} \, \sigma^a_i \sigma^b_{i+1}
\end{equation}
with 12 real coefficients $c_{ab}$\,, with Pauli operators $\sigma^i_a$\,, with $i$ as a site index, identifying $\sigma^{n+1}_a = \sigma^1_a$. That is, we take $S$ to be the space of translation-invariant, nearest-neighbor Hamiltonians on an $n$-qubit chain, up to an additive constant $c \, \mathds{1}$, with $\dim S = 12$.  Because $\dim SU(2) = 3$, we have 3 ``gauge" degrees of freedom, corresponding to local unitary operators in $SU(2)$, which act by conjugation on $S$, with the same local unitary acting on each qubit.  We want to find $W \subset S$ with $\dim W = \dim S - \dim SU(2) = 9$.  

First consider the terms $a_{01} \sigma^1+a_{02} \sigma^2+ a_{03} \sigma^3$. There is a local unitary that diagonalizes this sum, which is to say that the terms $a_{01} \sigma^1+a_{02} \sigma^2+ a_{03} \sigma^3$ are unitarily equivalent to the terms $a'_{03} \sigma^3$ for some $a'_{03}$.   We can then remove the terms $a_{01} \sigma^1+a_{02} \sigma^2$ from the gauge-fixed subspace $W \subset V$, because any $H \in S$ is equivalent to some $H' \in V$.  Acting by conjugation with diagonal matrices in $SU(2)$ will preserve the $a'_{03} \sigma^3$ term, and it will rotate the $a_{11}  \sigma^1_i \sigma^1_{i+1}$ and $a_{12}  \sigma^1_i \sigma^2_{i+1}$ terms between each other.  In particular, there exists some diagonal matrix in $SU(2)$ that acts by conjugation to rotate the two terms $a_{11}  \sigma^1_i \sigma^1_{i+1} + a_{12}  \sigma^1_i \sigma^2_{i+1}$ into the single term $a'_{11}  \sigma^1_i \sigma^1_{i+1}$.  Then any $H \in S$ is equivalent to some $H' \in V$, where $V$ is of the form of Eqn.\! ($\ref{eq:spinchain}$), but with $a_{01}=a_{02}=a_{12}=0$, using only 9 coefficients.  

From dimensional considerations, for almost all $H \in W$, $H$ will be isolated point in the intersection of $W$ with the orbit of $H$ under local unitary conjugation.  In other words, there are no infinitesimally nearby points in $w$ that are related by local unitaries.  However, the orbit of $H$ under local unitary conjugation may intersect $V$ in several isolated points.  (In keeping with the analogy to gauge theory, one might say that the gauge-fixing is not global, and there is ``Gribov ambiguity.")  Similarly, the complexification $W_\C \subset V_\C$ will contain at least one representative of each orbit of $V_\C$ under conjugation by 1-local invertible operators.  Then we can search for duals by optimizing
\begin{equation}
\underset{H \in W_\C}{\operatorname{arg min}}\, \| \textrm{spectrum}(H)-\textrm{spectrum}(H_0) \|
\end{equation}
rather than optimizing over $V_\C$.  

It is difficult to perform a global optimization numerically; that is, it is difficult to know that the minima found by the optimization algorithm are global minima and not just local minima.  Luckily, we know the value of the global minimum (zero), and we know at least one of the global minima (i.e., $H_0$), but we want to find \textit{all} global minima.  We perform a gradient-descent-type search over $W_\C$, starting with a random initial point $H_{init} \in W_\C$.  After finding a global minimum $H \in W_\C$ with $\| \textrm{spectrum}(H)-\textrm{spectrum}(H_0) \|=0$, we check whether it is related to $H_0$ by some combination of local unitary operations, reflections of the chain, or transposition.  (Despite searching over the ``gauge-fixed" subspace $V_\C$, we must still check whether $H$ is related to $H_0$ by local unitaries, due to the imperfect gauge-fixing discussed above.)    If $H$ is not related to $H_0$ in this way, it is not a true dual.

In practice, we consistently found that for a randomly generated $H_0$ and randomly generated starting point $H_{init}$, the search algorithm found either the global minimum $H_0$, or some other global minimum $H$ related to $H_0$ by local unitaries, reflections, or transpose.   Because the starting point $H_{init}$ was chosen independently of the original point $H_0$, and because several choices of random initial point consistently led to $H_0$ or some a related Hamiltonian, we expect that $H_0$ has no duals.  The more times the search is repeated with different starting points, the more confident one becomes that $H_0$ has no duals.  

One might worry that the global minimum $H_0$ sits in a wide basin, i.e.\! the function $\| \textrm{spectrum}(H)-\textrm{spectrum}(H_0) \|$ is flat around $H_0$.  Then $H_0$ would be found for most starting points, while perhaps the duals of $H_0$ are global minima that sit in narrower basins for some reason, making them difficult to find with a local search algorithm.  If this were the case, consistently finding $H_0$ from a random starting point would \textit{not} necessarily be evidence that $H_0$ has no duals.  On the other hand, we can repeat the search for many choices of $H_0$.  If the latter scenario were true, where $H_0$ sometimes has duals that sit in narrow basins, one would expect that some of these randomly generated $H_0$ actually sit in the narrower basin, in which case the search would find the dual, because the dual would then be in the wider basin. 

Given 1000 randomly generated $H_0 \in W_\C$, with a randomly generated starting point $H_{init}$ for each $H_0$, we found that in all of the trials, the search algorithm identified either the global minimum $H_0$ or another global minimum $H$ related by local unitaries, reflections of the chain, or transposition.  We therefore strongly believe that these $H_0$ have no duals, although we do not undertake a rigorous analysis of the efficacy of this probabilistic verification.  However, assuming that these randomly-generated $H_0$ indeed have no duals, we can apply Theorem \ref{thrm:constduals} to imply that almost all translation-invariant, nearest-neighbor Hamiltonians on 7-qubit spin chains have no duals.  Alternatively, we have demonstrated that for generic translation-invariant spin chains, the Hamiltonian may be uniquely determined from the spectrum.

\subsection{Discovering dualities}

So far, we have been focused on finding examples to establish that almost all Hamiltonians of a given class do not have duals.  However, suppose that we are given a spectrum that \textit{does} have duals.  Then the algorithms mentioned above can find the dual descriptions.

For concreteness, consider the one-dimensional Ising model, given in Eqn. (\ref{eq:ising}) in the introduction, repeated here:
\begin{equation*}
H = J \sum_{i=1}^{n-1} \sigma^z_i \sigma^z_{i+1} + h \sum_{i=1}^n \sigma^x_i\,.
\end{equation*}
In fact, the Ising model is special, and has many duals that are geometrically $2$-local plus boundary terms, even when we require that they must be translation-invariant.  We discovered many duals numerically by searching over the relevant class of Hamiltonians.

To find the particular dual in Eqn. (\ref{eq:isingdual}), namely
\begin{equation*}
H = J \sum_{i=1}^n \sigma^x_i + h \sum_{i=1}^{n-1} \sigma^z_i \sigma^z_{i+1} \, - J \, \sigma_1^x + h \, \sigma_n^z\,,
\end{equation*}
we limited our search to a more restricted subspace of Hamiltonians which contains our dual of interest.  This subspace is defined by the class of Hamiltonians of the form
\begin{equation}
\label{eq:arbitraryclass}
\sum_{i=1}^{n-1}\left(\sum_{p=1}^3 a_p \, \sigma_i^p \sigma_{i+1}^p + b_p \, \sigma_i^p\right) + \sum_{p=1}^3 (c_p\, \sigma_1^p + d_p \, \sigma_N^p)
\end{equation}
where the $\sum_{p=1}^3 (c_p\, \sigma_1^p + d_p \, \sigma_N^p)$ terms are possible boundary terms.  To be clear, unlike above where we searched over \textit{all} Hamiltonians of a particular locality class, here we are restricting the search to an essentially arbitrary, smaller `locality class' of Hamiltonians that was deliberately chosen to contain a particular dual.

Numerically searching over the class of Hamiltonians in Eqn. (\ref{eq:arbitraryclass}) indeed recovered $H$, and also found the desired dual in Eqn. (\ref{eq:isingdual}).  Thus, by using locality to guide our search in the space of Hamiltonians, we may search for dualities of a given system if we have reason to believe that such dualities exist.

\section{Generalizations of Tensor Product Structures} \label{sec:gauge}

So far, we have restricted our discussion to finite-dimensional bosonic systems, i.e. hardcore bosons or generalized spin systems.  One desired line of generalization is to consider infinite-dimensional systems, mentioned at the end of the section.  Another line of generalization is to consider theories which are qualitatively local, but for which the observables do not form a strict tensor product structure, at least in the sense defined by Section \ref{sec:tps}.

Examples of local theories without strict TPS's are fermionic theories and gauge theories.  In Section \ref{sec:tps}, a TPS on a Hilbert space $\HH$ was defined as a collection of of subalgebras $\{\A_i\}$, $\A_i \in L(\HH)$, such that $[\A_i, \A_j]=0$ for $i \neq j$, $\A_i \cap \A_j = \mathds{1}$, and $\bigvee_i \A_i = L(\HH)$.  Fermionic lattice theories  do not directly fit this description, because fermionic operators at different sites anti-commute.  One might therefore wonder in what sense fermionic theories ``local": is commutation necessary for locality?  In fact, commutation relations \textit{are} generally necessary to prevent signalling between distant locations.  But physical theories with fermions are nonetheless local, because the Hamiltonian contains terms with even products of nearby fermion operators, and these terms do commute with each other.  By restricting the algebra of observables on the Hilbert space to the subalgebra of ``physical" observables -- namely,  even products of single-fermion operators -- we can then arrange the physical algebra into mutually commuting subalgebras associated with spatial regions.  This general notion is captured by a ``net of observables," the basic structure used in algebraic quantum field theory \cite{haag}, and one can easily adapt the field-theoretic definition to discretized lattice systems.  We equip a Hilbert space with a set $S$ of spatial sites, like the points of a lattice.  Crucially, these sites do not correspond to tensor factors of the Hilbert space; they are just abstract labels.  Subsets  $U \subset S$ are ``regions," and we have

\noindent \textbf{Definition (Net of observables):} A net of observables on Hilbert space $\HH$ is a subalgebra of ``physical" observables $\A \subset L(\HH)$, along with a set of sites $S$, and an assignment of a subalgebra $\A(U) \subset \A$ to each region $U \subset S$.  The subalgebras must satisfy $\A(S)=\A$, along with
\begin{enumerate}
\item $\A(U) \subset \A(V)$ for $U \subset V$
\item $[\A(U),\A(V)]=0$ for disjoint regions $U \cap V = \emptyset$
\item $\A(U) \cap \A(V) = \mathds{1}$ for disjoint regions $U \cap V = \emptyset$
\end{enumerate}
\noindent Finally, one might also require:
\begin{enumerate}
\setcounter{enumi}{3}
\item The map $\A(U) \otimes \A(V) \to \A(U \cup V),\, A \otimes B \mapsto AB$ is injective.
\end{enumerate}

This definition is similar to that used by \cite{kitaev}. In the context of a net of observables, a local Hamiltonian would be one that may be written as a sum of terms in $\A(U)$ for small regions $U$, perhaps where $S$ has the additional structure of a geometric lattice. 

As a simple example of a net of observables, consider a Hilbert space $\HH = \bigotimes_{i=1}^n \HH_i$ with an ordinary TPS.  Then the naturally associated net of observables would be defined by $\A(U) = L(\HH_U)$ for $\HH_U = \bigotimes_{i \in U} \HH_I$.  However, the purpose of defining a net of observables is that not all nets must be associated with explicit tensor factorizations of $\HH$.  For instance, given a fermionic theory on a lattice, we would define $\A(U)$ to be generated by products of even numbers of fermion creation and annihilation operators.  Then $\A(S) \subsetneq  L(\HH)$, i.e. only fermion-even operators are considered physical, and the Hilbert space has no natural TPS.\footnote{In one dimension, local fermionic systems may be re-written as local bosonic systems using the Jordan-Wigner transformation, and the bosonic system has an ordinary TPS.  In that sense, one-dimensional fermionic systems do have natural TPS.  However, in higher dimensions, the immediate generalization of the Jordan-Wigner transformation does not map local Hamiltonians to local Hamiltonians, so higher-dimensional fermionic systems do not have natural TPS in which local Hamiltonians appear local with respect to the TPS. (There do exist constructions that embed the Hilbert space of any fermionic theory into a larger Hilbert space with explicit TPS, and these embeddings preserve locality by using auxiliary degrees of freedom \cite{kitaev}.)}

Like fermionic theories, gauge theories also lack an ordinary TPS, at least when restricting to the ``physical" Hilbert space.  But, like fermionic theories, the local structure of gauge theories is suitably generalized by using a net of observables instead.  For simplicity, consider two-dimensional $\mathbb{Z}_2$-lattice gauge theory.  The full, ``unphysical" Hilbert space is the tensor product of qubit degrees of freedom  living on the edges of a square lattice.   So the full Hilbert $\overline{\HH}$ space is endowed with a natural TPS.  However, the physical Hilbert $\HH \subsetneq \overline{\HH}$  is the proper subspace gauge-invariant physical states, and the physical observables $L(\HH)$ consist of gauge-invariant observables on $\overline{\HH}$, restricted to the gauge-invariant subspace.  In general, a subspace of a space with an explicit TPS will not inherit the TPS in any natural way, so the physical Hilbert space will not have a natural TPS.  On the other hand, we can construct a net of observables by defining $\A$ to be the algebra of gauge-invariant observables, with $\A(U)$ the algebra of gauge-invariant observables local to $U$.  Gauge theory does not have the property that $\A(U) \otimes \A(V) \cong \A(U \cup V)$ for disjoint regions $U \cap V = \emptyset$, which would be true for any theory with an ordinary TPS, showing that an ordinary TPS would not have sufficed to capture the local structure of the theory.

We have seen that nets of observables provide a generalized notion of TPS sufficient to capture the local structure of fermions and gauge theory.  Do the uniqueness results at the heart of this paper generalize to theories whose local structure is described by a net  of observables, rather than a strict TPS?  That is, given an abstract Hilbert space and Hamiltonian, we can ask whether there exists a net of observables on the Hilbert space such that the Hamiltonian is local.  And then, given that such a net exists, we can ask whether it is unique.  

Questions about nets are harder to tackle than the analogous questions about ordinary TPS's.  To see why, let us reconsider a nuance in the discussion of ordinary TPS's that we have not addressed. Given some local system of qubits, rather than simply asking whether the system has a dual using a different set of qubit degrees of freedom, we might ask whether there is a dual using qudits.  That is, we may want to consider duals that use different ``types" of TPS's.  Up to unitary equivalence, a TPS on a Hilbert space is just characterized by the list of dimensions of the subsystems, and these must multiply to the total dimension, so it is easy to characterize the ``types" of TPS's: qubits, or qutrits, or some combination, etc.  Meanwhile, there are many more types of nets.  Indeed, on a given Hilbert space, there are more possible nets of observables, even up to unitary equivalence.  The net corresponding to fermions is different than the net corresponding to gauge theory, because they use different sorts of algebras $\A(U)$, and one could construct nets that do not obviously correspond to bosons, fermions, or gauge theories.   The harder question then becomes: given some Hamiltonian $H$ that looks local using a given net, does $H$ have any duals that not only use different local degrees of freedom but also use a different type of net?

While the above questions are certainly difficult, the following observation suggests they may tractable.  We already know the types of TPS are easy to characterize, controlled by the dimension of the Hilbert space.  Given some fixed $\dim(\HH)$, the TPS cannot have too many subsystems, assuming each subsystem has dimension greater than one.  The types of nets are similarly controlled.  As long as one assumes that the algebra $\A(U)$ is non-trivial for any region $U$ larger than some fixed size, then condition (4) in the definition of a net ensures that $\dim(\HH)$ will be exponential in the number of sites.  So for given Hilbert space $\HH$, a net cannot be constructed with more sites than about $\log(\dim(\HH))$, offering some control on the types of nets allowed.

None of the results in this paper directly apply to infinite-dimensional systems.  There are three types of infinities to consider.  First, a theory with non-hardcore bosons will have infinite-dimensional Hilbert spaces at each lattice site.  Second, in the continuum limit, there are infinitely many lattice sites per fixed volume, associated with UV divergences.  Third, in the large system limit, there are infinitely many lattice sites at fixed spacing.  The large-system limit alone may yield interesting complications when attempting to reproduce the finite-dimensional results.  The discussion in Section \ref{sec:main} relies essentially on the well-behavedness of the map from a Hamiltonian to its spectrum.  However, in the large-system limit, the eigenvalues may vary non-analytically with respect to the Hamiltonian, leading famously to phase transitions.  Another result of non-analyticity is that properties which are true generically for finite-size systems may not be true generic infinite-size systems.  For instance, finite-size local Hamiltonians are generically non-degenerate; that is, a random local perturbation of a degenerate local Hamiltonian will break the degeneracy. But certain infinite-size lattice systems have topological order, with a ground state degeneracy that is robust to any local perturbation.  In particular, there exist open neighborhoods in the space of infinite two-dimensional lattice Hamiltonians such that all Hamiltonians in the neighborhood have degenerate spectrum.  Therefore, one cannot na\"{i}vely rule out the possibility that there exists a region of nonzero volume in the space of infinite-size local Hamiltonians where the Hamiltonians all have duals.  

\section{Discussion}

\subsection{Summary of results}

We began by formally defining a tensor product structure (TPS) on a Hilbert space, allowing one to pose clear questions about the existence of TPS's for which a Hamiltonian is local.   First we observed that for some fixed Hamiltonian $H$, questions about the existence of a TPS in which $H$ is local may be translated into questions about the existence of local Hamiltonians with the same spectrum as $H$.   With this perspective, we showed that almost all Hamiltonians do \textit{not} have any TPS for which the Hamiltonian is local.   Equivalently, generic Hamiltonians are not isospectral to any local Hamiltonian. 

On the other hand, physical systems are distinguished by the property that they \textit{are} local in some TPS, or at least approximately so.  We therefore considered Hamiltonians known to have some local TPS and argued that the local TPS is generically unique.  Equivalently, a generic local Hamiltonian is uniquely determined by its spectrum.  Put a third way, generic local Hamiltonians do not have ``duals."   The argument for this claim involves two parts: first, we proved that if there exists a single example of a local Hamiltonian without any duals, then almost all local Hamiltonians have no duals. Second, we found numerical examples of local Hamiltonians for small systems that do not have any duals, effectively proving that almost all Hamiltonians on these systems do not have any duals.  We speculated that these results may be extended to arbitrarily large finite-dimensional systems, with the possibility of interesting subtleties in the infinite-size limit.  

Finally, we presented a generalization of a TPS, suitable for fermions and gauge theories.  Further generalizations discussed below address situations where only a subspace of the full Hilbert space is equipped with a TPS, perhaps with relevance to the bulk side of holographic theories.

\subsection{The minimal data needed to understand a quantum system}

In this paper we argue that given the spectrum of a Hamiltonian that is local in some TPS, then generically the local TPS is uniquely determined.  In fact, one can also determine the local TPS by merely knowing the time evolution of a single generic state in the Hilbert space, without otherwise knowing $H$.  The time evolution of a generic state $|\Psi\rangle$ has the form
\begin{equation}
\label{typicalState1}
|\Psi(t)\rangle = \sum_j \alpha_j \, e^{- i E_j t} |E_j\,\rangle 
\end{equation}
which contains a non-zero amplitude $\alpha_j$ for each of the eigenstates $|E_j\rangle$ with energy $E_j$.  Taking the Fourier transform of $|\Psi(t)\rangle$ with respect to time, we can determine the spectrum and hence the local TPS.

However, even with a known TPS, much remains to be understood about the unitary evolution of states.  Much research is dedicated to the subject of expressing the wavefunction as a sum of decoherent classical branches \cite{Jess1, Jess2, Zurek1}.  This research generally assumes the existence of some underlying TPS and it has been recognized that it would be preferable to have the TPS emerge naturally in the same way as the branches \cite{Schlosshauer}; our results suggest a way to do that.  

\subsection{Geometry on the TPS}

Given both a low-energy state and a TPS, recent work \cite{XL1, Carroll1} suggests one can construct a metric on the discrete sites of the TPS.  The distance assigned between sites is dictated by the mutual information between the subsystems, giving a ``geometry" that depends on the state.  It would be possible to combine this approach with the work in this paper, determining both a TPS \textit{and} a notion of distance between subsystems,  starting from just the spectrum. First one determines the most local TPS, then finds the ground state of the Hamiltonian with respect to that TPS, and finally uses the mutual information of the ground state to define distances between subsystems.

In this paper, we already associate a graph to the TPS, based on which sites are directly interacting under the Hamiltonian. The graph approximately describes the topological structure of the space, while the proposal of \cite{Carroll1} would assign lengths to the edges of the graph, upgrading the topological data to geometric data.  However, when the Hamiltonian is already known, it may be more natural to rely on dynamical notions of distance like the light-cone or butterfly velocity, rather than asking about the mutual information of a state at fixed time.

\subsection{Quantum simulation} \label{sec:sim}

Recent progress has been made on the construction of universal quantum simulators \cite{Cubitt}.  In particular, consider a finite lattice system in $d$ spatial dimensions, governed by local Hamiltonian $H$ on Hilbert space $\HH$.  Then one can always construct a local, two-dimensional spin system with Hamiltonian $H'$ on Hilbert space $\HH'$, such that the low energy subspace of $H'$ reproduces the spectrum of $H$ with arbitrary precision.   

Because the simulator requires many auxiliary degrees of freedom, the number of lattice sites used in $\HH'$ will be larger than the number of sites present in the original system $\HH$, so $\HH \not\cong \HH'$ and the systems are not dual in the strict sense used above.  However, one might consider the notion of a TPS for a subspace of the full Hilbert space.  Restricting attention to the low energy subspace of the device, one could in principle find the TPS corresponding to the simulated system.  This situation may be analogous to the AdS/CFT duality in which the TPS of the bulk gravity theory only describes a subspace of the full Hilbert space.  This is further discussed in Section \ref{sec:QG} below. 

\subsection{Why locality?} \label{sec:anth}

Until now, we have avoided the question of why to prefer one TPS over another.  Instead, we have simply asserted that we are interested in TPS's for which dynamics appear local. If one treats the wavefunction and its Hamiltonian without any preferred basis as the only fundamental data of a quantum system, then \textit{a priori} all TPS's are equally valid descriptions of the system.\footnote{This is a radical view if taken literally.  If one built a quantum simulator of the kind discussed in Section \ref{sec:sim}, the radical view would suggest that the TPS of the simulated system has the same ontological status as the TPS of the simulation device.}

Because the world around us has local interactions, it is natural that we are interested in TPS's with local dynamics.  However, one might ask why our experience privileges a particular TPS for the universe -- namely, the TPS associated with spatial degrees of freedom?  One possible answer is that local interactions are an essential ingredient for localized observers.  For contrast, consider some randomly chosen TPS, in which interactions are non-local.  A hypothetical observer ``localized" in this TPS will quickly become delocalized, so perhaps observers in such a TPS cannot exist for extended periods.  Instead, only a TPS with local dynamics can naturally describe localized observers, and their experience will privilege that local TPS.

The existence of localized observers may also require more than just a local TPS.  For example, even local interactions may be strongly coupled and chaotic, such that localized objects quickly become maximally entangled with their environment.  One might expect that such dynamics do not allow localized observers, because such observers would quickly become delocalized despite having only local interactions.

A measure of entanglement growth was considered in \cite{piazza} as a criterion for choosing a TPS, though the author restricted the analysis to TPS's related by Bogoliubov transformations.  When searching for a TPS with slow entanglement growth, one must decide for which class of states to consider the entanglement.  One possibility is to consider random product states \cite{Carroll2}, while another natural choice would be low-energy states.

\subsection{Complexity}

The computational complexity of a unitary operator -- the number of local quantum gates needed to contstruct the operator -- is an important notion in quantum information theory and features in discussions of quantum gravity \cite{Scott1, Lenny1, Lenny2}.  The complexity of an operator depends crucially on the choice of TPS.  

Given a fixed TPS, random unitary operators will have complexity that is exponential in system size, as demonstrated by a dimension-counting argument.  Moreover, by an argument similar to that of Section \ref{sec:existence}, generic unitary operators will have no TPS in which they have low complexity.  

However, if a Hamiltonian is local in some TPS, the time-evolution operator $e^{-iHt}$ will have much smaller complexity in that TPS, at least for times sub-exponential in the system size.  Because the locality of $H$ determines the growth rate of complexity of $e^{-iHt}$, at least for sufficiently small times, an alternate description of the local TPS is the TPS in which $e^{-iHt}$ has minimal complexity at small times.

\subsection{Quantum gravity} \label{sec:QG}

The most well-known of the dualities in quantum gravity is the AdS/CFT correspondence between strongly coupled $\mathcal{N}\!=\!4$ super Yang-Mills in $3+1$--dimensions (the boundary theory) and weakly coupled quantum gravity in $\text{AdS}_5 \times S^5$ (the bulk theory) \cite{Juan1, Ed1}.  This duality is unlikely to satisfy the precise definition of duality used in this paper, even using the generalization of Section \ref{sec:gauge}. In particular, the TPS in the bulk is only defined for a subspace of states of the complete Hilbert space. These are states associated with small perturbations of the geometry around a flat AdS background \cite{EC1, EC2}.

However, when the state contains a black hole, for example, it does not make sense to talk about the same approximately-local degrees of freedom that existed in flat space. The discrepancy is especially manifest in tensor network toy models of AdS/CFT, where the model of a black hole involves tearing out tensors from the network \cite{EC2}. This model completely removes some of the bulk lattice sites, and instead the `correct' TPS for the subspace of states containing the black hole consists of the remaining bulk sites, together with new lattice sites at the boundary of the black hole. Describing different subsets of states in the Hilbert space with different TPS's in a coherent way seems to require yet another generalization tensor product structures.

The question of whether the boundary theory or bulk theory is ``more local" is somewhat subtle.  The bulk gravitational theory will necessarily have small non-local interactions, but it also has far fewer degrees of freedom at each ''lattice site" than does the boundary theory, where there is a large $N \times N$ matrix of operators associated to each site. The bulk TPS has a much smaller algebra of local operators, since it divides the Hilbert space up into much smaller subsystems. One might therefore describe the bulk TPS as more local than the boundary TPS when considering the low energy subspace, even though the Hamiltonian is only approximately local with respect to the bulk TPS.

More speculatively, we might also guess based on our arguments in Section \ref{sec:anth} that the boundary theory may be too strongly coupled and chaotic to describe localized observers.

\subsection{The SYK model}

A toy model for AdS/CFT, the Sachdev-Ye-Kitaev model \cite{SYK1, SYK2, SYK3}, is particularly relevant to the discussions in this paper.  The Hamiltonian of the theory is comprised of $N$ majorana fermions $\psi_a$ with all-to-all $4$-local coupling terms\footnote{Note that while the Hamiltonian is $4$-local, it is completely geometrically non-local, with every site interacting with every other site.}:
\begin{equation}
\label{SYKham}
H = \sum_{a<b<c<d} j_{abcd} \, \psi_a \psi_b \psi_c \psi_d
\end{equation}
The coefficients $j_{abcd}$ are sampled from i.i.d. random Gaussians, describing an ensemble of Hamiltonians. Since this ensemble is the fermionic analog of a class of bosonic local Hamiltonians considered in this paper, we might expect that generic SYK Hamiltonians would not have any local duals.  On the other hand, when one disorder-averages the SYK Hamiltonian and takes the expectation of observables over the probability distribution for $\{j_{abcd}\}$, one can remarkably rewrite the theory in terms of degrees of freedom that include a type of Einstein-Dilaton gravity in $1+1$ dimensions \cite{SYK3}.  As a consequence, one can compute the spectrum of the bulk gravity theory by computing the spectrum of the Majorana theory (in a particular limit), which is comparatively easier to treat \cite{SYK4, Verbaarschot1}.  

In accordance with our intuition, it is likely that the complete description of the dynamics is not even approximately local at scales smaller than the 1+1d AdS scale. Nonetheless, it is interesting that this alternative description is able to exist at all, when the Hamiltonian itself is generic within some class of local Hamiltonians.

\subsection{Final remarks}
There are many open questions about whether our results extend to the generalized notion of TPS suitable for fermions and gauge theories discussed in Section \ref{sec:gauge}, as well as to infinite-dimensional systems or approximately local TPS's.  Furthermore, while we have provided evidence that recovery of the TPS from the spectrum of spin chains is generically possible in principle, we have not discussed practical measures to determine whether a local TPS exists for a given spectrum or how to find it apart from numerically searching through possible TPS's. It appears that finding the most local TPS for a given spectrum is computationally impractical (using classical computation) for all but the smallest Hilbert spaces, but it is possible that there may be very good heuristic algorithms.  It would be interesting if there was, in contrast, an efficient quantum algorithm to find local TPS's.

\subsection*{Acknowledgements}
We would like to thank Daniel Bump, Dylan Butson, Emilio Cobanera, Benjamin Lim, Edward Mazenc, Xiao-Liang Qi, Semon Rezchikov, Leonard Susskind, Arnav Tripathy, Ravi Vakil, and Michael Walter for their valuable discussions and support.  We are also especially grateful to Patrick Hayden and Frances Kirwan for their valuable insights and feedback, and for reviewing this manuscript.  JC is supported by the Fannie and John Hertz Foundation and the Stanford Graduate Fellowship program.  DR is supported by the Stanford Graduate Fellowship program.

\end{document}